\documentstyle{article}

\newtheorem{theorem}{Theorem}

\newtheorem{lemma}[theorem]{Lemma}
\newtheorem{proposition}[theorem]{Proposition}
\newtheorem{remark}[theorem]{Remark}
\begin{document}
{\bf \Large  Entangled Markov Chains generated by Symmetric Channels}
\\
\\
\hfill {\bf 
\ Takayuki Miyadera
\footnote{Research Center for Information Security,
National Institute of Advanced Industrial 
Science and Technology (AIST), \\
Daibiru building 1102,
Sotokanda, Chiyoda-ku, Tokyo, 101-0021, Japan\\
e-mail: miyadera-takayuki@aist.go.jp}}
\\
\\
{\bf Abstract:}
A notion of entangled Markov chain was introduced by 
Accardi and Fidaleo in the context of quantum random walk.
They proved that, in the finite dimensional case, the corresponding states 
have vanishing entropy density, but they did not prove that they are entangled.

In the present note this entropy result is extended to the infinite 
dimensional case under the assumption of finite speed of hopping.

Then the entanglement problem is discussed for spin $1/2$, entangled Markov 
chains generated by a binary symmetric channel with 
hopping probability $1-q$. The von Neumann entropy 
of these states, restricted on a sublattice is explicitly 
calculated and shown to be independent of the size of 
the sublattice. This is a new, purely quantum, phenomenon.

Finally the entanglement property between the sublattices
${\cal A}(\{0,1,\ldots,N\})$ and ${\cal A}(\{N+1\})$ is investigated
using the PPT criterium. It turns out that, for $q\neq 0,1,\frac{1}{2}$ 
the states are non separable, thus truly entangled, 
while for $q=0,1,\frac{1}{2}$, they are separable. 
\\
\section{Introduction}
Motivated by recent developments in quantum information theory,
Accardi and Fidaleo introduced a Markov chain, called ``entangled",
and including a quantum version of classical random walks \cite{AccardiFidaleo}. 
They consider a quantum spin chain and impose the following conditions 
on its state.
\begin{itemize}
\item[(i)] It should be a quantum Markov chain \cite{Topics}. 
\item[(ii)] It should be purely generated \cite{FNW}. 
\item[(iii)] Its restriction on at least one maximal Abelian 
subalgebra, should be a classical random walk.
\item[(iv)] It should be uniquely determined, up to 
arbitrary phases, by its classical restriction.
\end{itemize}
In order to fix the notations, let us briefly review their definition.

A one-sided (or two sided) infinite quantum spin chain is defined by
a $1$--dimensional lattice, which in our case will be ${\bf N}$ (or ${\bf Z}$).
On each site of the lattice there is a spin degree of freedom. 

Its observables are represented by the algebra ${\bf B}({\cal H})$,
of all the bounded operators on some separable Hilbert space ${\cal H}$.
Typically (but not for random walks) its dimension is finite (say 
$d <\infty$). In this case ${\bf B}({\cal H})$ is a $d\times d$ matrix algebra,
${\cal A}(\{x\})\simeq M_d({\bf C}) (x\in {\bf N} \ (\mbox{or } {\bf Z}))$.
For each finite region $\Lambda$ in ${\bf N} \ (\mbox{or } {\bf Z})$, 
algebra of observables with respect to $\Lambda$ is defined 
by ${\cal A}(\Lambda):=\otimes_{x \in {\Lambda}}{\cal A}(\{x\})$.
The total algebra of observables is defined as the closure of their union,
\begin{eqnarray*}
{\cal A}:=\overline{\cup_{ \Lambda} {\cal A}(\Lambda)}^{\Vert\ \Vert},
\end{eqnarray*}
where the closure is taken with respect to norm topology.

On this chain we consider a class of states defined as follows.
First we will consider the case of one-sided chain (defined on ${\bf N}$). 
Then the state is extended to the two-sided chain by imposing translation invariance. 

To define an \textit{entangled Markov state}
\cite{AccardiFidaleo}, we begin with 
its finite volume version.
Suppose there exist a probability distribution $\{P(i)\}_{i\in \Omega}$
on a set $\Omega$ whose cardinality is same as the dimension of ${\cal H}$, 
and a transition probability 
$\{P(i\to j)\}_{(i,j)\in \Omega\times\Omega}$ 
from $\Omega$ to itself. We define a vector in 
${\cal H} \otimes  {\cal H} \otimes \cdots \otimes {\cal H}$ ($N+1$-times) 
by 
\begin{eqnarray*}
|\Psi_N\rangle:=\sum \sqrt{ P(i_0)P(i_0\to i_1)P(i_1 \to i_2) 
\cdots P(i_{N-1} \to i_N)}|i_0 i_1 \cdots i_N\rangle,
\end{eqnarray*} 
where $\{|i\rangle\}$ is an orthonormal basis in ${\cal H}$ and
$|i_0 i_1 \cdots i_N\rangle:= |i_0\rangle 
\otimes |i_1\rangle \otimes \cdots \otimes |i_N\rangle$.
It is easily checked that the norm of this vector is $1$ and 
therefore it defines a state over ${\cal A}(\{0,1,\ldots,N\})$ by
\begin{eqnarray*}
\omega_N(\cdot):=\langle\Psi_N|\cdot|\Psi_N\rangle.
\end{eqnarray*} 
Accardi and Fidaleo has shown that the infinite volume limit 
\begin{eqnarray*}
\omega(\cdot):=\lim_{N\to \infty}\omega_N(\cdot),
\end{eqnarray*}
exists and defines a state on ${\cal A}$ which is a quantum Markov 
chain in the sense of \cite{Topics}. The states in this class of quantum 
Markov chains are called \textit{entangled Markov states}.
In the infinite dimensional case ($\mbox{dim}{\cal H}=\infty$) 
a sub--class of these entangled states can be regarded as a quantum version 
of the classical random walks. The finite dimensional case 
(${\cal H}={\bf C}^d$) is also interesting in the context of 
statistical mechanics of spin chains.

In this note we estimate the entropy density of such states
when the dimension of ${\cal H}$ is infinite.

In the case $d=2$ and under the assumption that its generating 
classical transition probability is symmetric and stationary, we explicitly
compute the entropy of finite sub--lattices. Finally its entanglement property 
is examined.

\section{Entropy Density}

In this section we consider a one-sided spin chain and 
entangled Markov states over it. 
To estimate entropy of a finite sublattice, we compute the 
restriction of the chain on ${\cal A}(\{0,1,\ldots,N\})$. 
It is  easy to verify that the coefficients, in the given basis, 
of the density matrix of the chain, localized on $\{0,1,\ldots,N\}$, are:
\begin{eqnarray*}
&&\left(\rho_N\right)_{i_0i_1\cdots i_N,j_0 j_1\cdots j_N}
:=
\omega(|i_0 i_1 \cdots i_N\rangle \langle 
j_0 j_1 \cdots j_N| \otimes {\bf 1})
\\
&&=\sum \sqrt{P(i_0)P(i_0\to i_1)P(i_1 \to i_2) 
\cdots P(i_{N-1} \to i_N)P(i_{N} \to i)}
\\
&&
\sqrt{P(j_0)P(j_0\to j_1)P(j_1 \to j_2) 
\cdots P(j_{N-1} \to j_N)P(j_{N} \to i)}.
\end{eqnarray*}
The following theorem is easy to prove. 
\begin{theorem}
The restriction of an entangled Markov state on the Abelian subalgebra ${\cal M}$, 
generated by the matrices which are diagonal in the given basis, 
gives a classical Markov chain. 
If such a classical chain is stationary 
(i.e., if $\sum P(i)P(i\to j) =P(j)$ is satisfied 
for each $j$) its Shannon entropy density is (see e.g. \cite{CT})
$-\sum_i P(i)\sum P(i\to j)\log P(i\to j)$.
\end{theorem}
To estimate the von Neumann entropy of the density matrix $\rho_N$, 
the following lemma is crucial.
\begin{lemma}\label{lemma1}
For an arbitrary $N\in {\bf N}$ and for any $A \in {\cal A}(
\{0,1,\ldots,N\})$, 
\begin{eqnarray*}
\omega(A)=\langle \Psi^{N+1}|A|\Psi^{N+1}\rangle
\end{eqnarray*}
holds. That is, for strictly local operator, taking into account 
one additional 
site is sufficient.
\end{lemma}
 {\bf Proof:}
 For sufficiently large $M$, $\langle \Psi_{M}|
 A|\Psi_M\rangle$ can be expressed as
 \begin{eqnarray*}
&&\langle \Psi_M|A|\Psi_M\rangle
\\
&&=\sum \sqrt{P(i_0)P(i_0\to i_1)P(i_1 \to i_2) 
\cdots P(i_{N-1} \to i_N)P(i_{N} \to i_{N+1})}
\\
&&
\sqrt{P(j_0)P(j_0\to j_1)P(j_1 \to j_2) 
\cdots P(j_{N-1} \to j_N)P(j_{N} \to i_{N+1})}
\\
&&
P(i_{N+1} \to i_{N+2})\cdots P(i_{M-1}\to i_{M})
\langle i_0 i_1 \cdots i_N|A|j_0 j_1\cdots j_N\rangle.
\end{eqnarray*}
Thanks to $\sum_{i_{l}}P(i_{l-1} \to i_l)=1$, 
it does not depend on $M$ as soon as $M\geq {N+1}$.
\hfill Q.E.D.
\\
Thus the following theorem holds.
\begin{theorem}\label{theorem1}
Suppose the dimension $d$ of Hilbert space ${\cal H}$ is finite. 
For any $N\in {\bf N}$, von Neumann entropy of 
$\rho_N$ satisfies
\begin{eqnarray*}\label{theorem3}
S_{vN}(\rho_N):=-\mbox{tr}(\rho_N\log \rho_N)
\leq \log d.
\end{eqnarray*}
\end{theorem}
{\bf Proof:}
The previous lemma means that  
\begin{eqnarray*}
\rho_N =\mbox{tr}_{N+1}
\left(|\Psi_{N+1}\rangle \langle 
\Psi_{N+1}|\right)
\end{eqnarray*}
holds.
If we put $\sigma_{N+1}:=\mbox{tr}_{0,1,\ldots,N}
\left(|\Psi_{N+1}\rangle \langle 
\Psi_{N+1}|\right)$, according to Schmidt decomposition theorem\cite{Schmidt},
the purity of $|\Psi_{N+1}\rangle \langle \Psi_{N+1}|$ implies that 
the eigenvalues of $\rho_N$ coincide with ones of $\sigma_{N+1}$, and
\begin{eqnarray*}
S_{vN}(\rho_N)=S_{vN}(\sigma_{N+1})
\end{eqnarray*}
holds.
Since $\sigma_{N+1}$ is a state on ${\bf C}^d$,
its von Neumann entropy is bounded from above by 
$\log d$.
Thus we can conclude that 
for any $N$,
\begin{eqnarray*}
S_{vN}(\rho_N)\leq  \log d
\end{eqnarray*}
holds. 
\hfill Q.E.D.
\\
This allows to simplify the proof of the following result, obtained 
in \cite{AccardiFidaleo}.
\begin{proposition}
For finite $d$, any entangled Markov state has 
vanishing mean von Neumann entropy.
\end{proposition}
\begin{remark}
It is known \cite{FNW} that the vanishing of the mean entropy is 
not equivalent to the purity of the state. For instance in
the case $d=2$, $P(0)=P(1)=1$ and $P(0\to 0)=P(1\to 1)=1$,
the resulting state is an equal mixture of the two pure states, 
$|000\cdots\rangle\langle \cdots 000|$ and $|111\cdots\rangle\langle \cdots 111|$.
\end{remark}
For the infinite dimensional case we obtain the following result.
Consider the case $\Omega={\bf Z}$, which includes a quantum version of 
the classical random walks on a lattice.

Theorem \ref{theorem1} cannot be directly applied and in fact 
even for single site its von Neumann entropy can be infinite \cite{[OhyPet93]}. 
We, however, are interested in the case when initially 
the distribution is localized and it gradually expands 
to its neighbours. 
That is, typically the initial distribution $P(\cdot)$ has a 
compact support, say $\Lambda \subset {\bf Z}$. Moreover, the speed of 
hopping should be finite. That is, there exists $V<\infty$ 
such that for all $i \in {\bf Z}$,
\begin{eqnarray*}
\mbox{max}\{|c|: P(i \to i+c)\neq 0\}\leq V
\end{eqnarray*}
holds.
Under these conditions, the following theorem holds.
\begin{theorem}
For localized initial distributions and finite hopping range $V$, 
the von Neumann entropy of $\rho_N$ is bounded from above as follows:
\begin{eqnarray*}
S_{vN}(\rho_N)
\leq \log\left(|\Lambda|+2V(N+1) \right).
\end{eqnarray*}
\end{theorem}
{\bf Proof:}
The range of summation for
\begin{eqnarray*}
|\Psi_{N+1}\rangle:=\sum \sqrt{ P(i_0)P(i_0\to i_1)P(i_1 \to i_2) 
\cdots P(i_{N} \to i_{N+1})}|i_0,i_1,\ldots,i_{N+1}\rangle,
\end{eqnarray*} 
can be finite. 
As in the theorem \ref{theorem1}, density operators 
$\mbox{tr}_{N+1}|\Psi_{N+1}\rangle \langle \Psi_{N+1}|$ and 
$\mbox{tr}_{0,1,\ldots,N}(|\Psi_{N+1}\rangle \langle \Psi_{N+1}|)$
show the same value of von Neumann entropy.
Since 
\begin{eqnarray}
&&\mbox{tr}_{0,1,\ldots,N}(|\Psi_{N+1}\rangle 
\langle \Psi_{N+1}|)
=\sum_{i_{N+1}}\sum_{j_{N+1}}\sum_{i_N}
\tilde{P}(i_N) \nonumber \\
&&
\sqrt{P(i_N \to i_{N+1})}\sqrt{P(i_N \to j_{N+1})}
|i_{N+1}\rangle \langle j_{N+1}|,\label{eqneq}
\end{eqnarray}
holds, where $\tilde{P}(i_N)$ is defined as 
\begin{eqnarray*}
\tilde{P}(i_N):=\sum P(i_0)
P(i_0 \to i_1)P(i_1 \to i_2) 
\cdots P(i_{N-1}\to i_N).
\end{eqnarray*}
The summation for $i_{N+1}$ and $j_{N+1}$ in 
(\ref{eqneq}) runs over finite range whose 
cardinality is bounded by  $|\Lambda|+2V(N+1)$. 
\hfill Q.E.D.
\\
Thus we obtain the following.
\begin{theorem}
The von Neumann entropy density of an entangled Markov chain, with localized 
initial distributions and finite hopping range $V$, vanishes.
\end{theorem}

\section{$d=2$: Symmetric case}

In this section we analyze the simplest example of 
entangled Markov state, namely the case $d=2$ with symmetric
transition probability. That is, the state is generated by a channel:
\begin{eqnarray*}
P(0\to 0)&=&q
\\
P(1\to 1)&=&q,
\end{eqnarray*}
where $0\leq q\leq 1$ holds.
We, in addition, assume stationarity and 
thus for $q\neq 0,1$, 
\begin{eqnarray}
P(0)=P(1)=\frac{1}{2}
\label{equal}
\end{eqnarray}
must hold. For simplicity, also for $q=0,1$, 
we assume (\ref{equal}) holds.
 In this case Lemma \ref{lemma1} enables 
us to diagonalize the state $\rho_N$ for arbitrary $N$
as shown by the following theorem.
\begin{theorem}
For $q\neq \frac{1}{2}$, there are only two nonvanishing 
eigenvalues for $\rho_N$, and they are 
\begin{eqnarray*}
\lambda_+&:=&\frac{1}{2}+\sqrt{q(1-q)}
\\
\lambda-&:=&\frac{1}{2}-\sqrt{q(1-q)}.
\end{eqnarray*}
Their corresponding eigenvectors are 
respectively,
\begin{eqnarray*}
|\Psi_N^{\pm}\rangle
&:=&\frac{1}{2\sqrt{\frac{1}{2}\pm \sqrt{q(1-q)}}}
\sum_{i_0\cdots i_N}
\left(\sqrt{1-q}\right)^{\sum_{\alpha=1}^N (i_{\alpha-1}\oplus i_{\alpha})}
\left(\sqrt{q}\right)^{N-\sum_{\alpha=1}^N (i_{\alpha-1}\oplus i_{\alpha})}
\\
&&
\left(
\sqrt{(1-q)^{i_N}q^{1-i_{N}}}
\pm
\sqrt{(1-q)^{1-i_N}q^{i_{N}}}
\right)
|i_0 i_1 \cdots i_N\rangle,
\end{eqnarray*}
where $\oplus$ means summation with mod $2$ (XOR operation).
($0$ is its eigenvalue with $2^{N+1}-2$ multiplicity.)
For $q=\frac{1}{2}$, 
$\rho_N$ is a pure state over ${\cal A}(\{0,1,\ldots,N\})$.
\end{theorem}
{\bf Proof:}
By lemma \ref{lemma1}, $\rho_N$ is equal to
$\rho_N=\mbox{tr}_{N+1} |\Psi_{N+1}\rangle \langle \Psi_{N+1}|$. 
This fact and Schmidt decomposition theorem \cite{Schmidt}
 shows that 
$|\psi_{N+1}\rangle$
can be expressed as 
\begin{eqnarray*}
|\Psi_{N+1}\rangle
=\sum_{l} \sqrt{\lambda_l}|\Psi_N^{l}\rangle
\otimes |e_l\rangle,
\end{eqnarray*} 
where $\lambda_l (l=\pm)$ are the common eigenvalues of $\rho_N$ and 
$\sigma_{N+1}:=\mbox{tr}_{1,2,\ldots,N}|\Psi_{N+1}\rangle
\langle \Psi_{N+1}|$ and $\{|e_l\rangle\}$'s 
are the eigenvectors of $\sigma_{N+1}$.
Thus to obtain $\lambda_l$ and $|\Psi_N^{l}\rangle$,
we should first diagonalize the $2\times 2$ matrix $\sigma_{N+1}$
which can be easily computed to be:
\begin{eqnarray*}
\sigma_{N+1}
=\frac{1}{2}\left(|0\rangle \langle 0|
+|1\rangle \langle 1|\right)
+\sqrt{q(1-q)}\left(|0\rangle \langle 1|
+|1\rangle \langle 0|\right).
\end{eqnarray*} 
Its eigenvalues are 
\begin{eqnarray*}
\lambda_{\pm}
=\frac{1}{2}\pm \sqrt{q(1-q)}
\end{eqnarray*}
and the corresponding eigenvectors are
\begin{eqnarray*}
|e_{\pm}\rangle
:=\frac{1}{\sqrt{2}}\left(|0\rangle 
\pm |1\rangle
\right).
\end{eqnarray*}
The vectors
$\sqrt{\lambda_{\pm}} |\Psi_N^{\pm}\rangle \otimes |e_{\pm}\rangle$
are obtained by applying ${\bf 1}\otimes |e_{\pm}\rangle\langle e_{\pm}|$
to $|\psi_{N+1}\rangle$. This gives:
\begin{eqnarray*}
|\Psi_N^{\pm}\rangle
=&&\frac{1}{\sqrt{\frac{1}{2}\pm\sqrt{q(1-q)}}}
\sum_{i_0 \cdots i_N}
\sqrt{P(i_0)P(i_0 \to i_1)
\cdots P(i_{N-1} i_N)}
\\&&
\frac{1}{\sqrt{2}}
\left(\sqrt{P(i_N\to 0)}\pm
\sqrt{P(i_N\to 1)}
\right)|i_0 \cdots i_N\rangle,
\end{eqnarray*}
which is directly deformed into 
the desired form.\\
In case of $q=\frac{1}{2}$, one of the above eigenvalues $\lambda_-$ vanishes 
and $\rho_N$ is shown to be pure.
\hfill Q.E.D.
\\
In view of the above theorem and the Schmidt decomposition 
theorem\cite{Schmidt},
the following theorem is
obvious.
\begin{theorem}
For $q\neq \frac{1}{2}$, 
von Neumann entropy of $\rho_N$ for any $N\in {\bf N}$ is
\begin{eqnarray*}
S_{vN}(\rho_N) = S_{vN}(\sigma_{N+1})
&=&-\left(\frac{1}{2}+\sqrt{q(1-q)}\right)
\log \left(\frac{1}{2}+\sqrt{q(1-q)}\right)
\\
&&
- \left(\frac{1}{2}-\sqrt{q(1-q)}\right)
\log \left(\frac{1}{2}-\sqrt{q(1-q)}\right).
\end{eqnarray*}
For $q=\frac{1}{2}$, $S_{vN}(\rho_N)=0$ for any $N \in {\bf N}$.
\end{theorem}
\begin{remark}
It is not difficult to verify that the above technique 
can be used for a general (non symmetric) channel with $d=2$.
\end{remark}
Now we investigate the entanglement property of the 
states. Let us consider an entangled Markov state 
generated by a symmetric channel and 
its restriction to sublattice ${\cal A}(\{0,1,\ldots,N,N+1\})$ 
which is written as $\rho_{N+1}$ in a density matrix.
If we divide the sublattice ${\cal A}(\{0,1,\ldots,N,N+1\})$ into 
${\cal A}(\{0,1,\ldots,N\})$ and ${\cal A}(\{N+1\})$, 
is the state $\rho_{N+1}$ separable or entangled between 
them?
The following theorem gives the answer.
\begin{theorem}
For $q\neq 0,1,\frac{1}{2}$, the above defined 
$\rho_{N+1}$ is entangled (i.e., inseparable) 
between ${\cal A}(\{0,1,\ldots,N\})$ and ${\cal A}(\{N+1\})$.
For $q=0,1,\frac{1}{2}$, $\rho_{N+1}$ is separable.
\end{theorem}
{\bf Proof:}
Let us consider the two dimensional subspace spanned by 
$|\Psi_N^{+}\rangle$ and $|\Psi_N^{-}\rangle$. 
and its (normalized but not orthogonal) basis:
\begin{eqnarray*}
|\Phi_N(0)\rangle
&:=&\sqrt{\lambda_+}|\Psi_N^+\rangle
+\sqrt{\lambda_-}|\Psi_N^-\rangle
\\
&=&
\sum_{i_0 \cdots i_N}\sqrt{P(i_0 \to i_1)
P(i_1 \to i_2) \cdots P(i_N \to 0)}
|i_0 i_1 \cdots i_N\rangle
\end{eqnarray*}
\begin{eqnarray*}
|\Phi_N(1)\rangle
&:=&\sqrt{\lambda_+}|\Psi_N^+\rangle
-\sqrt{\lambda_-}|\Psi_N^-\rangle
\\
&=&\sum_{i_0 \cdots i_N}\sqrt{P(i_0 \to i_1)
P(i_1 \to i_2) \cdots P(i_N \to 1)}
|i_0 i_1 \cdots i_N\rangle.
\end{eqnarray*}
It is easy to see that the following relations hold,
\begin{eqnarray*}
\langle \Phi_N(0)|\Phi_N(0)\rangle
&=&\langle \Phi_N(1)|\Phi_N(1)\rangle=1
\\
\langle \Phi_N(0)|\Phi_N(1)\rangle &=&\lambda_+-\lambda_-.
\end{eqnarray*}

The expansion of $|\Psi_{N+1}^{\pm}\rangle$ in this basis is: 
\begin{eqnarray*}
|\Psi_{N+1}^{\pm}\rangle
=\frac{1}{2\sqrt{\lambda_{\pm}}}
\left(\sqrt{q}\pm \sqrt{1-q}
\right)\left(
|\Phi_N(0)\rangle \otimes |0\rangle
\pm |\Phi_N(1)\rangle \otimes |1\rangle
\right).
\end{eqnarray*}
Therefore the density matrix $\rho_{N+1}$ can be written as 
\begin{eqnarray*}
\rho_{N+1}
&=&\lambda_+ |\Psi_{N+1}^+ \rangle
\langle \Psi_{N+1}^+ |
+\lambda_- |\Psi_{N+1}^-\rangle
\langle \Psi_{N+1}^- |
\\
&=&
\frac{1}{2}
\left(
|\Phi_N(0)\rangle \langle \Phi_N(0)|
\otimes |0\rangle \langle 0|
+|\Phi_N(1)\rangle \langle \Phi_N(1)|
\otimes |1\rangle \langle 1|
\right)
\\
&&
+\frac{\lambda_+ -\lambda_-}{2}
\left(
|\Phi_N(0)\rangle \langle \Phi_N(1)|
\otimes |0\rangle \langle 1|
+
|\Phi_N(1)\rangle \langle \Phi_N(0)|
\otimes |1\rangle \langle 0|
\right)
\end{eqnarray*}
Since it is can be identified with a matrix in ${\bf C}^2
\otimes {\bf C}^2$, the PPT (positive partial transpose) 
criterion of \cite{Peres,Horodeckis} can be used to check 
its separability\cite{lowrank}. 
According to this criterion $\rho_{N+1}$ is separable
if and only if the partially transposed matrix 
\begin{eqnarray*}
\rho^{PT}_{N+1}
&:=&
\frac{1}{2}
\left(
|\Phi_N(0)\rangle \langle \Phi_N(0)|
\otimes |0\rangle \langle 0|
+|\Phi_N(1)\rangle \langle \Phi_N(1)|
\otimes |1\rangle \langle 1|
\right)
\\
&&
+\frac{\lambda_+ -\lambda_-}{2}
\left(
|\Phi_N(0)\rangle \langle \Phi_N(1)|
\otimes |1\rangle \langle 0|
+
|\Phi_N(1)\rangle \langle \Phi_N(0)|
\otimes |0\rangle \langle 1|
\right).
\end{eqnarray*}
is still positive. Let us prove that, in the present case, 
$\rho^{PT}_{N+1}$ is not positive in general. In fact,
computing $\langle \varphi|\rho^{PT}_{N+1}|\varphi\rangle$  
where $\varphi$ is the normalized vector:
\begin{eqnarray*}
|\varphi\rangle
:=\frac{1}{\sqrt{2}}
\left(|\Phi_N(1)\rangle \otimes |0\rangle
-|\Phi_N(0)\rangle \otimes |1\rangle 
\right)
\end{eqnarray*}
gives
 \begin{eqnarray*}
 \langle \varphi|\rho^{PT}_{N+1}|\varphi\rangle
 =2\sqrt{q(1-q)}
 \left(
 \sqrt{q(1-q)}
 -\frac{1}{2}
 \right)
 \end{eqnarray*}
 which is negative when $q\neq 0,1,\frac{1}{2}$.
 Thus we can conclude that $\rho_{N+1}$ is 
 an entangled state between ${\cal A}(\{0,1,\ldots,N\})$
 and ${\cal A}(\{N+1\})$.
 \\
 For $q=\frac{1}{2}$, by the previous lemma 
it is easily seen that
 $\rho_{N+1}$ is just a product state of 
 pure states.
 \\
 For $q=1$, a straightforward calculation shows that 
 $\rho_{N+1}$ has the form, 
 \begin{eqnarray*}
 \rho_{N+1}=\frac{1}{2}\left(|00\cdots00\rangle
 \langle 00\cdots00|
 +|11\cdots11\rangle \langle 11\cdots 11|\right)
 \end{eqnarray*}
 which is obviously separable.
 \\
 For $q=0$, $\rho_{N+1}$ can be written as,
 \begin{eqnarray*}
 \rho_{N+1}=\frac{1}{2}\left(|01\cdots 01\rangle
 \langle 01 \cdots 01|
 +|10 \cdots10 \rangle \langle 10 \cdots 10 |\right)
 \end{eqnarray*}
(here we assumed $N$ is odd).
It also is obviously separable. 
 \hfill Q.E.D.


\end{document}